\documentclass[11pt]{article}
\usepackage{amssymb}
\usepackage{amsfonts}
\usepackage{graphics}
\def\rlx{\relax\leavevmode}
\def\inbar{\vrule height1.5ex width.4pt depth0pt}
\def\IZ{\rlx\hbox{\small \sf Z\kern-.4em Z}}
\def\IR{\rlx\hbox{\rm I\kern-.18em R}}
\def\ID{\rlx\hbox{\rm I\kern-.18em D}}
\def\IC{\rlx\hbox{\,$\inbar\kern-.3em{\rm C}$}}
\def\IN{\rlx\hbox{\rm I\kern-.18em N}}
\def\one{\hbox{{1}\kern-.25em\hbox{l}}}
\def\beq{\begin{equation}}
\def\eeq{\end{equation}}
\def\bea{\begin{eqnarray}}
\def\eea{\end{eqnarray}}
\def\ber{\begin{array}}
\def\eer{\end{array}}

\begin{document}

\begin{titlepage}

July 2000 \hfill{UTAS-PHYS-99-20}\\
\mbox{}\hfill {ADP-00-05/T393}\\
\mbox{}\hfill{physics/0001066}
\vskip 1.6in
\begin{center}
{\Large {\bf The genetic code as a periodic table:  }}\\[5pt]
{\Large  {\bf algebraic
aspects }}\\[5pt]
\end{center}

\normalsize
\vskip .4in

\begin{center}
J. D. Bashford \\
{\it Centre for the Structure of Subatomic Matter, University of Adelaide} \\
{\it Adelaide SA 5005, Australia} \\
\par \vskip .1in \noindent
P. D. Jarvis \\
{\it School of Mathematics and Physics, University of Tasmania}\\
{\it GPO Box 252-21, Hobart Tas 7001, Australia }\\

\end{center}
\par \vskip .3in

\begin{center}
{\Large {\bf Abstract}}\\
\end{center}

The systematics of indices of physico-chemical properties of codons
and amino acids across the genetic code are examined. Using a simple
numerical labelling scheme for nucleic acid bases, $A=(-1,0)$,
$C=(0,-1)$, $G=(0,1)$, $U=(1,0)$, data can be fitted as low order
polynomials of the 6 coordinates in the 64-dimensional codon weight
space. The work confirms and extends the recent studies by Siemion et al. 
(1995) (BioSystems 36, 63-69) of the conformational parameters. Fundamental 
patterns in the data such as codon periodicities, and related harmonics and 
reflection symmetries, are here associated with the structure of the set of 
basis
monomials chosen for fitting. Results are plotted using the Siemion one-step
mutation ring scheme, and variants thereof. The connections between the
present work,
and recent studies of the genetic code structure using dynamical symmetry
algebras, are pointed out.
 
%\vspace{1cm}
%\begin{center}
%{\bf Physics and Astronomy Classifications}. 02.10.
%\end{center}
\vspace{1cm}

\end{titlepage}

\section{Introduction and main results}
Fundamental understanding of the origin and
evolution of the genetic code (Osawa et al., 1991) must be grounded in
detailed knowledge
of the intimate relationship between the
molecular biochemistry of protein synthesis, and the retrieval
from the nucleic acids of the proteins' stored design information.
However, as pointed out by Lacey and Mullins in 1983, although
`the nature of an evolutionary biochemical edifice must reflect \ldots
its constituents, \ldots properties which
were important to prebiotic origins may not be of relevance to
contemporary systems'. Ever since the final elucidation
of the genetic code, this conviction has led to many studies
of the basic building blocks themselves, the amino acids and the nucleic acid
bases. Such studies have sought to catalogue and understand the
spectrum of physico-chemical characteristics of these molecules, and
of their mutual correlations. The present work is a contribution to
this programme.

Considerations of protein structure
point to the fundamental importance of amino acid
hydrophilicity and polarity in determining folding and enzymatic capability,
and early work (Woese et al., 1966; Volkenstein, 1966; Grantham, 1974)
concentrated on these aspects; Weber and Lacey (1978) 
extended the work to mono- and di-nucleosides. Jungck (1978)
concluded from a compilation of more than a dozen properties that
correlations between amino acids and their corresponding anticodon
dinucleosides were strongest on the scale of hydrophobicity/hydrophilicity
or of molecular volume/polarity. For a comprehensive review
see Lacey and Mullins (1983).

Subsequent to this early work, using statistical sequence
information, conformational indices of amino acids in protein
structure have been added to the data sets (Goodman and Moore, 1977).
Recently Siemion (1994a, 1994b) has considered the 
behaviour of these parameters across the genetic code,
and has identified certain periodicities and pseudosymmetries
present when the data is plotted in a certain rank ordering
called `one-step mutation rings', being generated by a
hierarchy of cyclic alternation of triplet base letters (Siemion and 
Stefanowicz,1992). 
The highest level in this hierarchy is the alternation of the second base letter,
giving three major cycles based on the families $U$, $C$ and $A$, each sharing
parts of the $G$ family. The importance of the second base in relation to
amino acid
hydrophilicity is in fact well known (Weber and Lacey, 1978;
Lacey and Mullins, 1983; Taylor and Coates, 1989), and the existence of three
independent correlates of amino acid properties, again associated
with the $U$, $C$ and $A$ families, has also been statistically established
by principal component analysis (Sj\"{o}strom and W\"{o}ld, 1985).

Given the existence of identifiable patterns in the genetic code in
this sense, it is of some interest and potential importance to
attempt to describe them more quantitatively. Steps along these lines
were taken in 1995 by Siemion et al. With a linear rank ordering of amino
acids according to `mutation angle' $\pi k/32$, $k= 1, \ldots, 64$ along
the one-step mutation rings, parameter $P^{\alpha}$ was reasonably
approximated by trigonometric functions which captured the essential
fluctuations in the data.

The metaphor of a quantity such as the Siemion number $k$, allowing the 
genetic code to be arranged in a way which best reflects its structure, in
analogy with elemental atomic number $Z$ and the
chemical periodic table, is an extremely powerful one.
The present paper takes up this idea, but in a more flexible way which
does not rely on a single parameter. Instead, a natural labelling
scheme is used which is directly related to the combinatorial
fact of the triplet base codon structure of the genetic code,
and the four-letter base alphabet. Indeed,
any bipartite labelling system which identifies each of the four
bases $A, C, G, U$, extends naturally to a composite labelling
for codons, and hence amino acids. We choose for bases two coordinates
as $A =(-1,0)$, $C=(0,-1)$, $G=(0,1)$, $U=(1,0)$, so that codons are
labelled as ordered sextuplets, for example $ACG = (-1,0,0,-1,0,1)$.

In quantitative terms, any numerical indices of amino acid or codon properties,
of physico-chemical or biological nature, can then be modelled as
some functions of the coordinates of this codon `weight space'.
Because of other algebraic approaches to the
structure of the genetic code, we take polynomial functions (for simplicity,
of as low order as possible).
This restriction
does not at all exclude the possibility of periodicities and
associated symmetry patterns in the data.
In fact, as each of the six coordinates takes discrete values
$0, \pm 1$, appropriately chosen monomials can easily reproduce such
effects (with coefficients to be fitted which reflect the relative
strengths of various different `Fourier' components). 
Quite simply, the directness of the linear rank ordering, as given by
the Siemion number $k$, which suggests Fourier series analysis of
the data, is here replaced by a more involved labelling system,
but with numerical data modelled as simple polynomial functions.

The main results of our analysis are as follows.
In \S \ref{sec:codons} the labelling scheme for
nucleic acid bases is introduced, leading to $4^{N}$-dimensional
`weight spaces' for length-$N$ RNA strands: in particular,
$16$-dimensional for $N=2$, and $64$-dimensional for the sextuplet codon
labelling ($N$=3).
For $N=2$ the dinucleoside hydrophilicity, dinucleotide hydrophobicity, and 
free energy of formation of 2-base RNA duplices are considered.
Displayed as linear plots (or bar charts) on a ranking from 1 to 16, the
data have obvious
symmetry properties, and corresponding basis monomials are identified,
resulting in good fits. Only four coordinates are involved for these 16-
part data sets (see table \ref{table:dinucprops}). 
Moving in \S \ref{sec:conf} on to codon properties as correlated
to those of amino acids, Siemion number $k$ which establishes
amino acid ranking by mutation angle is briefly reviewed.
It is shown that the trigonometric approximation of Siemion et al. 
(Siemion et al., 1995) to the Chou-Fasman conformational parameters 
$P^{\alpha}$, $P^{\beta}$ (Chou and Fasman, 1975; Fasman, 1989) 
is effectively a four-parameter function which allows for periodicities of 
32/5, 8, 32/3 and 64 codons. Again, simple basis monomials having the required
elements of the symmetry structure of $P^{\alpha}$ are identified,
leading to a reasonable (four-parameter) fit. Results are displayed
as Siemion mutation-angle plots.  $P^{\beta}$ is treated in a
similar fashion. The method established
in  \S \S \ref{sec:codons} and \ref{sec:conf} is then applied in \S
\ref{sec:other} to other amino acid properties, including relative
hydrophilicity (Weber and Lacey, 1978) and Grantham polarity (Grantham, 1974).
It is clearly shown that appropriate polynomial functions can be fitted
to most of them (amino acid data is summarised in table
\ref{table:aaprops}).

In \S \ref{sec:concl} some concluding remarks, and outlook for
further development of these ideas are given. It is emphasised that,
while the idiosyncracies of real biology make it inappropriate to regard
this type of approach as anything but approximate,
nonetheless there may be some merit in a more rigorous
follow up to establish our conclusions in a statistically valid way.
This is particularly interesting in view of the appendix, \S
\ref{sec:dynsymm}.
This gives a brief review of algebraic work based on methods of
dynamical symmetries in the analysis of the excitation spectra
of complex systems (such as atoms, nuclei and molecules), which has
recently been proposed to explain the origin and evolution of the
genetic code. Specifically, it is shown how the labelling scheme
adopted in the paper arises naturally in the context of
models, based either on the
Lie superalgebra $A_{5,0}\sim sl(6/1)$, or the Lie algebra
$B_{6} \sim so(13)$, or related semisimple algebras.
The origins and nature of
the polynomial functions adopted in the paper, and
generalisations of these, are also discussed in the algebraic context.
The relationship of the present paper to the
dynamical symmetry approach is also sketched in \S \ref{sec:concl} below.

\section{Codon systematics}%check spelling in abstract etc
\label{sec:codons}

Ultimately our approach involves a symmetry between the
4 heterocyclic bases {U,C,A,G} commonly occurring in RNA.
A logical starting point then, is to consider the physical properties
of small RNA molecules.
Dinucleosides and dinucleotides in particular are relevant in the
informational context of the genetic code and anticode, and moreover
are the building blocks for larger nucleic acids (NA's). What follows in
this section is a numerical study of some properties of NA's consisting
of 2 bases, while in later sections NA's with 3 bases (i.e. codons and
anticodons) are considered in the context of the genetic code as being
correlated with properties of amino acids.

As mentioned in the introduction, we give each NA base coordinates in a
two-dimensional `weight space', namely $A=(-1,0)$, $C=(0,-1)$,
$G=(0,1)$, $U=(1,0)$ with the axes labelled $d,m$ respectively
\footnote{The choice $(\pm 1, \pm 1)$ and $(\pm 1, \mp 1)$ for the four
bases simply represents a 45$^\circ$ rotation of the adopted scheme, which
turns out to be more convenient for our purposes. The nonzero labels at
each of the four base positions are given by the mnemonic
`\textit{d}ia\textit{m}on\textit{d}'.  }. Dinucleosides and dinucleotides
are therefore associated with four coordinates $(d_1, m_1, d_2, m_2)$,
\textit{e.g.}
$AC= (-1,0,0,-1)$ with subscripts referring to the first and second base
positions.

The physical properties of nucleic acids we choose to fit to are the relative
hydrophilicities $R_{f}$ of the 16 dinucleoside monophosphates as obtained
by Weber and Lacey (1978), the relative hydrophobicity
$R_{x}$ of dinucleotides as calculated by Jungck (Jungck, 1978) from the
mononucleotide data of Garel et al. (1973), and the 16
canonical (Crick-Watson) base-pair stacking parameters of
Xia et al. (1998) used to compute the free heat of 
formation $ G^{0}_{37}$ of formation of duplex RNA strands at $37^{\circ}$ Centigrade.

It should be noted that the dinucleoside quantity $R_{x}$ was computed as 
the product of experimentally derived $R_{x}$ values for mononucleotides 
under the assumption that this determines the true dinucleotide values to 
within 95 \%. A result of this is that $R_{x}$ is automatically the same for 
dinucleotides $5'-XY-3'$ and $5'-YX-3'$,
(naturally the same holds for molecules with the reverse orientation);
thus $R_{x}$ is at best an approximate symmetry.

The 16 Turner free-energy parameters are a subset of a larger number
of empirically determined thermodynamic ``rules of thumb'' (see 
Xia et al.,1998; Mathews et al. 1999, for the most recent results)
, developed to predict free heats of formation of larger RNA and DNA 
molecules. The possibility that these rules have an underlying 
group-theoretical structure is a consideration for a future
work. For now it suffices to observe that due to geometry 
(see table \ref{table:dinucprops}) the duplex formed by
$5'-XY-3'$ with $3'-\bar{Y}\bar{X}-5'$ is just a rotated version of the duplex
$5'-\bar{X}\bar{Y}-3'$ with $3'-YX-5'$. Here $\bar{X}$ denotes the 
Crick-Watson complementary base to $X$. Furthermore the duplices formed by 
so-called ``self-complementary'' dinucleosides ($5'-GC-3'$ with $5'-CG-3'$ 
and $5'-AU-3'$ with $5'-UA-3'$) are thermodynamically suppressed due to the 
extra rotational symmetry (i.e there are 2 ways such a duplex can form) and 
one needs to include extra monomials to compensate for this.
It is easy to see that the above rotational symmetry corresponds
to the change of coordinates
\begin{equation}
(d_{1},m_{1},d_{2},m_{2}) \to (-d_{2},-m_{2},-d_{1},-m_{1})
\label{coorch}
\end{equation}
and thus we need look only at monomials which respect this symmetry,
for example $m_{1}-m_{2}$, $d_{1}d_{2}$ and $(d_{1}m_{2} + d_{2}m_{1})$.
The least-squares fit to the most recent values of the Turner parameters
(Xia et al., 1998) is shown in figure \ref{heatfit} and is given by
\begin{eqnarray}
 -G^{0}_{37}(d_{1},m_{1},d_{2},m_{2})&=& 1.133 + 0.02 ((m_{1}+d_{1})m_{2} + d_{2}(m_{1}-d_{1})) + 1.001 (m_{1}^{2} + m_{2}^{2}) \nonumber \\
& & \mbox{} -0.1 d_{1}d_{2}+ 0.035 (d_{1}-d_{2}) +0.1(m_{1}-m_{2}) \nonumber\\ 
& & \mbox{} +0.165 m_{1}m_{2}(m_{2}-m_{1}+1)+ 0.0225 d_{1}d_{2}(d_{2}-d_{1}-1).
\label{newgfit}
\end{eqnarray}
Here we have considered all linear and quadratic terms respecting the
symmetry Eq.(\ref{coorch}) and added cubic symmetry-breaking terms
which are specific to the self-complementary duplices. The number of 
monomials may be reduced with the identities:
\begin{eqnarray}
d_{i}^{2} &=& 1 - m_{i}^2 \\
m_{i} d_{i} & = & 0, 
\end{eqnarray}
Encouraged by this success we attempt a similar fit to $R_{f}$
and $R_{x}$. While there is no obvious underlying symmetry {\it a priori}
as in the previous discussion, one might expect these properties to be 
anti-correlated and so the same set of monomials is considered for each.
Qualitatively faithful fits may be obtained using a small number of monomials,
as shown in figure \ref{hydrofits}. The functions
\begin{eqnarray}
R_{f}(d{1},m_{1},d_{2},m_{2})&=&
0.191 - 0.087 (d^{2}_{1}+d^{2}_{2}) + 0.09 d_{1} +
0.107 d_{2} - 0.053 m_{1} - 0.077 m_{2}, \label{rffit}\\
R_{x}(d_{1},m_{1},d_{2},m_{2})&=&
0.3278 + 0.093 (d^{2}_{1} + d^{2}_{2}) - 0.1814 (d_{1} + d_{2}) + 0.0539
(m_{1} + m_{2}) \label{rxfit}
\end{eqnarray}
are seen to compare favourably to the experimental values and 
that moreover $R_{f}$ and $R_{x}$ are roughly anti-correlated.
Thus fitting to these using the same set of monomials seems to be a 
valid procedure in this initial approximation.

\section{Amino acid conformational parameters}
\label{sec:conf}

As a case study for amino acid properties (as
opposed to their correlated codon properties in \S \ref{sec:codons}
above) we consider the structural conformational parameters
$P^{\alpha}$ and $P^{\beta}$, which have been
discussed by Siemion (1994a, 1994b).
In 1995 Siemion et al. introduced a quantity $k$ , $k=1,\ldots, 64$, which
defined the so-called `mutation angle' $\pi k/32$ for a particular
assignment of codons (and hence of amino acids) in rank ordering. This is
a ramification of the four-ring ordering used above for plots
(expanded from 16 to 64 points), and
arises from a certain hierarchy of one-step base mutations. It
assigns the following $k$ values to the $NN'Y$ and $NN'R$ codons\footnote{
Individual codons are labelled so that these $Y$,$R$ positions are at
the midpoints of their respective $k$ intervals. Thus $GGR$ occupies
$0 \le k \le 2$, with nominal $k=1$ and codons $k(GGA)=0.5, \;
k(GGC)=1.5$.}

\begin{center}
	$
\begin{array}{cccccccc}
		1 & 3 & 5 & 7 & 9 & 11 & 13 & 15 \\
		
		GGY & GGR & GAR & GAY & AAY & AAR & CAR & CAY \\
		
		\hline
		
		17 & 19 & 21 & 23 & 25 & 27 & 29 & 31 \\
		
		UAY & UAR & UGR & UGY & UCY & UCR & GCR & GCY \\
		
		\hline
		
		33 & 35 & 37 & 39 & 41 & 43 & 45 & 47 \\
		
		ACY & ACR & CCR & CCY & CGY & CGR & CUR & CUY \\
		
		\hline
		
	   49 & 51 & 53 & 55 & 57 & 59 & 61 & 63 \\
		
	   UUY & UUR & GUR & GUY & AUY & AUR & AGR & AGY \\
	
	    \hline
		
	\end{array}
        $
\end{center}
wherein (as in the `four ring' scheme) the third base alternates as
$\ldots - G,A - U,C - C,U - A,G - \ldots$ for purine-pyrimidine occurrences
$\ldots -R-Y-Y-R- \ldots$ . This `mutation ring' ordering
corresponds to a particular trajectory around the diamond-shaped
representation of the genetic code (figure \ref{diamonds}), which is
pictured in figure \ref{simrings} (Siemion, 1994a) where nodes have been
labelled by amino acids.

Inspecting the trends of assigned $P^{\alpha}$ values for the amino
acids ordered in this way, a suggestive 8-codon periodicity, and a
plausible additional $C_{2}$ rotation axis about a spot in the centre of the
diagram, have be identified (Siemion, 1994a,1994b). Figure \ref{palphtrigfits}
gives various fits to this
data, as follows. Firstly, consideration of the modulation of the peaks
and troughs of the period-8 component, on either side of the centre
at $k=0$, leads to a trigonometric function (Siemion et al., 1995).
\begin{equation}
	P^{\alpha}_{S}(k) = 1.0 -[0.32+0.12\cos(\frac {k \pi}{16})] \cos(\frac{k
	\pi}{4}) -0.09 \sin(\frac{k \pi}{32})
	\label{PS}
\end{equation}
where the parameters are estimated simply from the degree
of variation in their heights (and $0.44=0.32 + 0.12$ is the average
amplitude). Least-squares fitting of the same data in fact leads to
a similar function,
\begin{equation}
	P^{\alpha}_{L}(k) = 1.02 -[0.22+0.21\cos(\frac {k \pi}{16})] \cos(\frac{k
	\pi}{4}) +0.005 \sin(\frac{k \pi}{32}).
	\label{L}
\end{equation}
From the point of view of Fourier series, however, the amplitude
modulation of the codon period-8 term in $P^{\alpha}_{S}$ or
$P^{\alpha}_{L}$
merely serves to add extra beats of period 32/5
and 32/3 of equal weight 0.06; an alternative might then be to allow
different coefficients. This gives instead the fitted function
\begin{equation}
	P^{\alpha}_{F}(k) = 1.02 -0.22 \cos(\frac{k \pi}{4})
	-0.11\cos(\frac { 3 k \pi}{16}) - 0.076 \cos(\frac{5 k \pi}{16})
	\label{F}
\end{equation}
which has no $\sin(\frac{k \pi}{32})$ term, but is almost
indistinguishable from equation (\ref{L}) above (note that
$0.22+0.21 \simeq 0.22+0.11+0.07 \simeq 0.32+0.12 = .44$). In figure
\ref{palphtrigfits} the $P^{\alpha}$ data
is displayed as a histogram along with $P^{\alpha}_{S}$, and
$P^{\alpha}_{F}$
above; as can be seen, both fits show similar trends,
and both have difficulty in reproducing the data around the
first position codons of the $C$ family in the centre of the diagram
(see caption to figure \ref{palphtrigfits}).

Basing the systematics of the genetic code
on numerical base labels, as advocated in the present
work, a similar analysis to the above
trigonometric functions is straighforward, but now in terms of
polynomials over the six codon (\textit{i.e.} trinucleotide)
coordinates $(d_{1}, m_{1}, d{_2}, m_{2}, d_{3}, m_{3})$. There is no
difficulty in establishing basic 8-codon periodic functions;
combinations such as $\frac{3}{2}d_{3}-\frac{1}{2}m_{3}$ (with values
$-\frac{3}{2}, -\frac{1}{2}, \frac{1}{2}, +\frac{3}{2}$ on $A, G, C,
U$), or more simply the perfect $Y/R$ discriminator $d_{3}-m_{3}$ (with values
$-1,+1$
on $R$,~$Y$ respectively) can be assumed. Similarly, terms such as
$d_{1} \pm m_{1}$ have period 16, and $d_{2} \pm m_{2}$ have period 64.
The required modulation of
the 8-codon periods can also be regained by including in the basis
functions for fitting a term such as
$d_{2}^{2}$, and finally an enhancement of the $C$ ring
family boxes $GCN$, $CCN$ is provided by the cubic
term $m_{1} m_2 (m_2-1)$. The resulting least-squares fitted function is
\begin{eqnarray}
\lefteqn{P^{\alpha}_{6}(d_{1},m_{1},d_{2},m_{2},d_{3},m_{3}) =} \nonumber \\
  && 0.86 +0.24 d_{2}^{2} + 0.21 m_{1} m_{2}(m_{2}-1) - 0.02 (d_{3} - m_{3})
	-0.075 d_{2}^{2}(d_{3} - m_{3}) \label{amon}
\end{eqnarray}
and is plotted against the $P^{\alpha}$ data in figure \ref{palphmonfits}.
The resulting fit\footnote{
In contrast to the trigonometric fits which are only intended to
fit the data for specified codons (indicated by the dots in figure
\ref{palphtrigfits}), the least-squares fit is applied for the
polynomial functions to all 64 data points. See Siemion (Siemion et al.,1995)
 and the captions to figures \ref{palphtrigfits} and \ref{palphmonfits}}
is rather insensitive to the
weights of $d_{3}$ and $m_{3}$ (allowing unconstrained coefficients
in fact results in identical weights $\pm .02$ for the linear terms
and $-.064, +.085$ for the $d_{2}^{2}$ coefficients respectively).
It should be noted that, despite much greater fidelity in the $C$ ring,
$P^{\alpha}_{6}$ shows
similar features to the least squares trigonometric fits $P^{\alpha}_{L}$
and $P^{\alpha}_{F}$
in reproducing the 8-codon periodicity less clearly than $P^{\alpha}_{S}$
(see figure \ref{palphtrigfits}). This indicates either that the
minimisation is fairly shallow at the fitted functions
(as suggested by the fact that $P^{\alpha}_{L}$ and $P^{\alpha}_{F}$
differ by less than $\pm 0.01$ over one period),
or that a different minimisation algorithm might yield somewhat
different solutions. To show the possible range of acceptable fits,
a second monomial is displayed in figure \ref{palphmonfits} whose
$d_{2}^{2}(d_{3} - m_{3})$ coefficient is chosen as $-0.2$ rather than $-0.075$.
This function plays the role of the original estimate $P^{\alpha}_{S}$
of figure \ref{palphtrigfits} in displaying a much more pronounced 
eight-codon periodicity than allowed by the least-squares algorithm.

The nature of the eight-codon periodicity is related to the
modulation of the conformational status of the amino acids through the
$R$ or $Y$ nature of their third codon base (Siemion and Siemion, 1994). A
sharper discriminator of this is the
difference $P^{\alpha}-P^{\beta}$, which suggests that a more appropriate 
basis for identifying numerical trends is with
$P^{\alpha} - P^{\beta}$ (the helix-forming potential) and
$P^{\alpha} + P^{\beta}$ (generic-structure-forming potential).
Although we have not analysed the data in this way, this
is indirectly borne out by separate fitting (along the same lines as
above) of $P^{\beta}$, for which \textit{no} significant component of
$(d_{3} - m_{3})$ is found. A typical five-parameter fit,
independent of third base coordinate, is given by
\begin{eqnarray}
\lefteqn{ P^{\beta}_{6}(d_{1},m_{1},d_{2},m_{2},d_{3},m_{3}) =} \nonumber \\
&& 1.02 + .26 d_{2}+ .09 d_{1}^{2} - .19 d_{2}(d_{1} - m_{1})
	  -.1 d_{1} m_{2}(m_{2}-1) - .16 m_{1}^{2} m_{2}(m_{2}-1).
\label{bmon}
\end{eqnarray}
Figure \ref{pbetafit} shows that this function
does indeed average over the third base $Y/R$ fluctuations
evident in the $A$ family data. A major component
appears to be the dependence on $(d_{1} - m_{1})$,
that is, on the $Y/R$ nature of the {\it first} codon base,
responsible for the major peaks and troughs visible on the $A$ and $U$
rings (and reflected in the $ d_{2}(d_{1} - m_{1})$ term). The cubic
and quartic terms follow the modulation of the data on the $C$ ring.

The suggested pseudosymmetries of the conformational parameters
are important for trigonometric functions of the
mutation angle, and for polynomial fits serve to identify
leading monomial terms with simple
properties. The $ d_{2}(d_{1} - m_{1})$ term in the fit of
$P^{\beta}_{6}$ above has been noted already in this connection.
In the case of $P^{\alpha}$, it should be noted that an
offset of 2 codons in the position of a possible $C_{2}$ rotation axis
(from $k=34$, between $ACY$ and $ACR$ to $k=32$, after $GCY$) changes
the axis from a pseudosymmetry axis (minima coincide with maxima
after rotation) to
a true symmetry axis (as the alignment of minima and maxima is shifted
by four codons), necessitating fitting by a period-eight component
which is \textit{even} about $k=32$. At the same time the large amplitude
changes in the $C$ ring appear to require an \textit{odd} function, and are
insensitive to whether the $C_{2}$ axis is chosen at $k=32$ or $k=34$.
The terms in $P^{\alpha}_{6}$ above have just these properties.

\section{Other amino acid properties}
\label{sec:other}

In this section we move from the biologically-measured conformational
parameters
to biochemical indices of amino acid properties. Two of the most
significant of these are the Grantham
polarity (Grantham, 1974) and the relative hydrophilicity as obtained by
Weber and Lacey (1978). Variations in
chemical reactivity have been considered (Siemion and Stefanowicz, 1992), but 
are not modelled here.

The composite Grantham index incorporates weightings for molecular
volume and molecular weight, amongst other ingredients (Grantham, 1974). 
From figure \ref{grafit} it is evident that
a major pattern is a broad 16-codon periodicity (indicative of a
term linear in $d_{2}$). 
Additional smaller fluctuations coincide approximately with the 8-codon
periodicity of the $Y/R$ nature of the third base ($d_{3}-m_{3}$ dependence).
Although there is much complex variation due to the first base, in
the interests of simplicity, the following fitted function ignores this
latter structure, and provides an approximate (2-parameter) model (see
figure \ref{grafit}):
\begin{equation}
	G_{6}(d_{1},m_{1},d_{2},m_{2},d_{3},m_{3}) =
	8.298 - 2.716 d_{2} - 0.14 (d_{3}-m_{3}).
\label{gmon}
\end{equation}
The pattern of amino acid hydrophilicity is also seen to possess an 8-codon 
periodicity. The 4-parameter fitted function considered, which is plotted 
in fig. \ref{aarffit}, is:
\begin{eqnarray}
\lefteqn{ R_{f6}(d_{1},m_{1},d_{2},m_{2},d_{3},m_{3}) = } \nonumber \\
 && 0.816 - 0.038 d_{2} - 0.043 m_{2} + 0.022 (d_{3} - m_{3}) + 
0.034 (1 - d_{2}) d_{2} (d_{3} - m_{3})
\label{rfmon}
\end{eqnarray}
As with the case of Grantham polarity, the 8-period extrema might
be more 'in phase' with the data if codons were weighted according to
usage, after the approach of Siemion (Siemion et al.,1995).

\section{Conclusions and outlook}
\label{sec:concl}

In this paper we have studied codon and amino acid correlations across the
genetic code starting from the simplest algebraic labelling scheme
for nucleic acid bases (and hence RNA or DNA strands more
generally). The relationship between the rank ordering of amino acids
according to Siemion number $k$, $k=1, ...,64$, and a description
of codons based on 3 dichotomic labels, had been established (figures
\ref{diamonds},\ref{simrings}).
In \S \ref{sec:codons} several dinucleoside properties
have been fitted as quadratic polynomials of the labels, and \S
\ref{sec:conf} and \S \ref{sec:other} have considered amino acid
parameters as correlated to codons (trinucleotides), namely
conformational parameters, Grantham polarity and hydrophilicity.
The types of data considered for fitting in our approach include
strictly physical information (amino acid molecular weight and volume),
physico-chemical indices (for example, the semi-empirical indicators of
dinucleoside free energy of formation, and the composite amino acid 
Grantham polarity), as well as biological measures (such as the 
conformational parameters, which are logarithmic measures
of amino acid usage in structural protein elements). 
As pointed out in the introductory discussion, {\it all} of these measures
should be considered as important aspects in the `optimisation' of the
genetic code (see also the remarks in the appendix, \S \ref{sec:dynsymm})
In all cases acceptable algebraic fitting is possible, and various
patterns and periodicities in the data are readily traced to the
contribution of specific monomials in the least-squares fit.

As pointed out in the appendix, \S \ref{sec:dynsymm}, our algebraic
approach is a special case of more general dynamical symmetry schemes
in which measurable attributes $H$ are given as combinations of
Casimir invariants of certain chains of embedded Lie algebras and
superalgebras (Bashford et al. 1997, 1998; Hornos and Hornos, 1993;
Schlesinger et. al 1998; Schlesinger and Kent, 1999; Forger et al., 1997).
The identification by Jungck (1978) of two or three major characters, to 
which all other properties are strongly
correlated, would similarly in the algebraic description mean the existence of
two or three distinct, `master' Hamiltonians $H_{1}$, $H_{2}$, $H_{3}$,
$\ldots$
(possibly with differing branching chains). In themselves these could
be abstract and need not have a physical interpretation, but all other
properties should be highly correlated to them,
\begin{equation}
	K = \alpha_{1}H_{1}+ \alpha_{2}H_{2} + \alpha_{3}H_{3}.
	\label{eq:masterH}
\end{equation}

Much has been made of the famous redundancy of the code in providing
a key to a group-theoretical description (Hornos and Hornos, 1993;
Forger et al. 1997). In the present framework
(see also Bashford et al., 1997; 1998), codon degeneracies take
second place to major features such as periodicity and other
systematic trends across the genetic code. Thus for example the noted 8-codon
periodicity of the conformational parameter $P^{\alpha}$ allows the
$Y$ codons for $k=25$, $UCY$, and $k=63$, $AGY$ both to be consistent
with {\it ser} (as the property attains any given value twice per 8-codon 
period, at $Y/R$ box $k=24+1=25$, and again 4 periods
later at the alternative phase $k= 56 + 7 =63$).

A related theme is the reconstruction of plausible ancestral codes
based on biochemical and genetic indications of the evolutionary
youth of certain parts of the existing code. For example the
anomalous features of arginine, {\it arg} which suggests that it is an
`intruder'
has led (Jukes, 1973) to the proposal of a more ancient code using
ornithine {\it orn} instead. This has been supported by the
trigonometric fit to $P^{\alpha}$ (Siemion et al., 1995; Siemion and 
Stefanowicz, 1996), as the inferred
parameters for {\it orn} actually match the fitted function better
than {\it arg} at the $k=43$, $k=61$ $CGR$, $AGR$ codons. Such variations
could obviously have some influence on the polynomial fitting, but at
the present stage have not been implemented\footnote{
The polynomial fits are to {\it all 64}
codons, not just those with greatest usage. In fact there is no
particular difficulty with {\it arg} in the $P^{\alpha}_{6}$ function
(see figure \ref{palphmonfits}).}.

To the extent that the present analysis has been successful in
suggesting the viability of an algebraic approach, further work with
the intention of establishing (\ref{eq:masterH}) in a statistically
reliable fashion may be warranted. What is certainly lacking to date
is any microscopic justification for the application of the
techniques of dynamical symmetry algebras (but see Bashford et al., 1997,1998).
However, it can be considered that in the path to the genetic code, the
primitive evolving and self-organising system of information storage
and directed molecular synthesis has been subjected to
`optimisation' (whether through error minimisation, energy
expenditure, parsimony with raw materials, or several such factors).
If furthermore the `space' of possible codes has the correct topology
(compact and convex in some appropriate sense), then it is not implausible that
extremal solutions, and possibly the present code, are associated with
special symmetries. It is to support the identification of such algebraic
structures that the present analysis is directed.

After this work was completed, we received a paper (Frappat et al.,2000) 
which gives a similar analysis of dinucleotide properties and correlations 
between physical-chemical properties of amino acids and codons based on a
particular algebraic scheme (see also Frappat et al., 1998). 
It should be emphasized that comparisons of such analyses based merely on 
the number of fitted parameters is not particularly illustrative at this
preliminary stage. One could modify, for example, Eq.(\ref{newgfit}) by cubic 
or quadratic transformations with the intent to minimise the number of 
parameters, but our motivation is to employ physical symmetry properties in 
an intuitive way. The number of parameters reflects the fact that our analysis
has no prior commitment to any given abstract algebraic scheme. Indeed 
reproducing the fitted $G^{0}_{37}$ of (Frappat et al., 2000) requires a 
judicous choice of cubic monomial terms.

\subsection*{Acknowledgements}
The authors would like to thank Elizabeth Chelkowska for assistance with
Mathematica (\copyright Wolfram Research Inc) with which the
least squares fitting was performed, and Ignacy Siemion for
correspondence in the course of the work.

\pagebreak

\renewcommand{\thesection}{\Alph{section}}
\setcounter{section}{0}
\section{Appendix: Dynamical symmetry algebras and genetic code structure}
\label{sec:dynsymm}

The radical proposal of Hornos and
Hornos (1993) to elucidate the genetic code structure using the
methods of dynamical symmetry algebras drew attention to the relationship
of certain symmetry-breaking chains in the Lie algebra $C_{3} \sim Sp(6)$ to 
the fundamental degeneracy patterns of the 64 codons. This theme has been
taken up subsequently using various different Lie
algebras (Schlesinger et al. 1998; Schlesinger and Kent, 1999; Forger et al.,1997) and also Lie superalgebras
(Bashford et al., 1997,1998,  Forger and Sachse,1998; Sorba and Sciarrino, 
1998).
In addition to possible insights into the code redundancy, a
representation-theoretical description also leads to a code
elaboration picture whereby evolutionary primitive, degenerate
assignments of many codons to a few amino acids and larger symmetry
algebras gave place, after symmetry breaking to subalgebras,
to the incorporation of more amino acids, each with fewer redundant codons.

In (Bashford et al., 1997,1998, Jarvis and Bashford, 1998)
emphasis was given not to the patterns of codon redundancy as such,
but rather to biochemical factors which have been recognised as
fundamental keys to be incorporated in any account of evolution from
a primitive coding system to the present universal one. Among these
factors is the primacy of the second base letter over the first and
third in correlating with such basic amino acid properties as
hydrophilicity (Woese et al., 1966 ;Volkenstein, 1966). Also, the
partial purine/pyrimidine dependence of the amino acid
assignments within a family box further underlines the informational
content of the third codon base (Siemion and Siemion, 1994) and necessitates a symmetry
description which distinguishes the third base letter.
In (Bashford et al., 1998) the amino acid degeneracy was
replaced by the weaker condition of anticodon degeneracy, leading to a
Lie superalgebra classification scheme using chains of subalgebras
of $A_{5,0} \sim sl(6/1)$ (see below for details).

A concomitant of any representation-theoretical description of the
genetic code is the `weight diagram' mapping the 64 codons to
points of the weight lattice (whose dimension is the rank of the algebra
chosen). Reciprocally, the line of reasoning advocated above and
applied in (Bashford et al., 1998) to the case of Lie
superalgebras suggests that \textit{any description using dynamical
symmetry algebras must be compatible with the combinatorial fact of
the four-letter alphabet, three-letter word structure of the code}.
The viewpoint adopted in the present paper is to explore the
implications of generic labelling schemes of this type, independently
of the particular choice of algebra or superalgebra. In particular,
as pointed out in \S \ref{sec:codons} above, the weight diagram is supposed
to arise from labelling each of the
three base letters of the codon alphabet with a pair of dichotomic
variables. Thus the only technical structural requirement for
Lie algebras and superalgebras compatible with the present work is the
existence of a 6-dimensional maximal abelian (Cartan) subalgebra,
and of 64-dimensional irreducible representations whose weight diagram
has the geometry of a six-dimensional hypercube in the weight lattice.
(The relationship between the base alphabet and the $Z_{2}\times
Z_{2}$ Klein four-group has been discussed by Bertman and Jungck
(1979).
As examples of a Lie algebra and a Lie superalgebra with this
structure, we here take the case of $B_{6} \sim SO(13)$ and
$A_{5,0} \sim sl(6/1)$ respectively (other examples would be
$SO(4)^{3}, sl(2/1)^{3}$).

The orthogonal algebra $SO(14)$ has been
suggested (Schlesinger et al., 1998) as a unifying scheme for variants of
the $Sp(6)$ models (Hornos and Hornos,1993; Forger et al.,1997). However,
from the present perspective, it is sufficient to take
spinor representations of the rank-6 odd orthogonal algebra $SO(13)$
which have dimension 64. Consider the subalgebra chain
\begin{eqnarray*}
	SO_{13} & \supset & SO^{(2)}_{4} \times SO_{9}  \\
	 & \supset &  SO^{(2)}_{4} \times SO^{(1)}_{4}  \times SO^{(3)}_{5};\\
	 SO^{(3)}_{5} & \supset & SO^{(3)}_{3} \times SO^{(3)}_{2}, \quad \mbox{or} \\
	SO^{(3)}_{5} \sim Sp^{(3)}_{4} & \supset & Sp^{(3)}_{2} \times Sp^{(3)'}_{2} ,
\end{eqnarray*}
where superscripts indicate base letter. The 64-dimensional
representation splits into 4 16-plets at the first breaking stage (the
four families labelled by second codon base letter, the latter being
distinguished as a spinor $(\frac 12, 0) + (0, \frac 12)$ of
$SO^{(2)}_{4}$).
The same pattern repeats for the first codon base $SO^{(1)}_{4}$
providing a complete labelling of the 16 family boxes (fixed first
and second base letter). The last stage gives two possible
alternatives for the third base symmetry breaking: in the first, each
family box would split into two doublets $(\frac 12, \frac 12) + (\frac 12,
-\frac 12)$ of $SO^{(3)}_{3} \times SO^{(3)}_{2}$, corresponding to a perfect
32-amino-acid-code $ 4 \rightarrow 2+2$, or to $Y/R$ degeneracy in anticodon
usage; in the second case, breaking of $Sp^{(3)'}_{2}$ to $U_{1}^{(3)}$
yields a family box assignment $2 \times (\frac 12, 0) + (0, + \frac 12)
+ (0, - \frac 12)$ coinciding with a 48-amino-acid-code, $4 \rightarrow
2+1+1$, or to perfect $Y$-degeneracy and $R$-splitting in amino acid
{\it usage}. In the eukaryotic code, the $4 \rightarrow 2+1+1$ 
family-box-pattern of anticodon usage is seen, whereas in the vertebrate
mitochondrial code, only partial $ 4 \rightarrow 2+2$ family box
splitting of anticodon usage is found (see below). Finally, the above
labels are all (up to normalisation) of the form $(0, \pm 1)$ or
$(\pm 1, 0)$ for each base letter (or $(\pm1, \pm 1)$ for the third
base for one branching) showing that this group-theoretical
scheme does indeed give a hypercubic geometry for the codon weight
diagram.

The $sl(6/1)$ superalgebra was advocated in a survey of possible Lie
superalgebras relevant to the genetic code (Bashford et al. 1997,1998),
and possesses irreducible, typical representations of dimension
$64$ which share many of the properties of spinor
representations of orthogonal Lie algebras (in the family $sl_{n/1}$ of
Lie superalgebras this class of representations has dimension $2^{n}$)
and so can be compared with spinors of the even- and odd-dimensional Lie
algebras of
rank $n$, namely $SO_{2n}$ and $SO_{2n+1}$ respectively). The
superalgebra branching chain related to the $SO(13)$ chain described
above is
\begin{eqnarray*}
	sl_{6/1} & \supset & sl^{(2)}_{2} \times sl_{4/1}  \\
     & 	\supset & sl^{(2)}_{2} \times sl^{(1)}_{2} \times  sl^{(3)}_{2/1};  \\
	sl^{(3)}_{2/1} & \supset & sl_{1/1} \quad \mbox{or}  \\
	sl^{(3)}_{2/1} & \supset & sl_{2} \times U_{1}
\end{eqnarray*}
where the last two steps correspond as above either to family box breaking
to $Y/R$ doublets
(as in many of the anticodon assignments of the vertebrate
mitochondrial code) or to a $4 \rightarrow 2+1+1$ pattern (as in the
anticodons of the eukaryotic code). The nature of the weight diagram
follows from knowledge of the branching in each of the above
embeddings. In fact both in the decomposition of the irreducible 64 to
families of 16, and in that of the 16 to family boxes of 4, there are a
doublet and two singlets of the accompanying $sl^{2}_{2}$ and
$sl^{2}_{1}$ algebras, so that the diagonal Cartan element (magnetic
quantum number) has the spectrum $0, \pm \frac 12$. A second diagonal
label arises because there is also an additional commuting $U_{1}$ generator
at each stage with value $ \pm 1$ on the two singlets and $0$ on the
doublet. Alternatively, the additional label may be taken as the
$\pm 1$ or $0$ shift in the noninteger Dynkin label of the commuting
$sl_{n/1}$ algebra ($n=4$ and $n=2$ respectively). Similar
considerations apply to the last branching
stage (Bashford et al., 1998), so that again the weight
diagram has the hypercubic geometry assumed
in the text of the paper.

In the dynamical symmetry algebra approach to problems of complex
spectra, important physical quantities such as the energy levels of
the system, and the transition probabilities for decays, are modelled
as matrix elements of certain operators belonging to the Lie algebra
or superalgebra. In particular, the Hamiltonian operator which
determines the energy is assumed to be a linear combination of a
set of invariants of a chain of subalgebras $G \supset G_{1} \supset G_{2}
\supset
\cdots T$:
$$
H = c_{1} \Gamma_{1} + c_{2} \Gamma_{2} + \cdots + c_{T} \Gamma_{T}
$$
for coefficients $c_{i}$ to be determined.
For states in a certain representation of the algebra $G$,
the energy can often be evaluated once the hierarchy of representations
of $\supset G_{1} \supset G_{2} \supset \cdots T$ to
which they belong is identified, as the invariants are functions of
the corresponding representation labels.

As has been emphasised above, the discussion of fitting of codon and amino acid
properties in the main body of the paper is independent of specific choices
of Lie algebras or
superalgebras. In fact, the polynomial functions of the 6 codon
coordinates may simply be regarded as \textit{generalised} invariants of
the smallest subalgebra common to all cases, namely the 6-dimensional Cartan
(maximal abelian)
subalgebra $T$ (so that there are several nonzero coefficients
$c_{T}$, with all other $c_{i}$ zero).
This approach is thus complementary to
detailed applications of a chosen symmetry algebra, where
the coefficients $c_{i}$ (including $c_{T}$) might accompany a specific
set of $\Gamma_{i}$ (functions of the whole hierarchy of labels, whose form
is fixed, depending on the subalgebra). However, because the weight
labels used in the present work already provide an unambiguous
identification of the 64 states, such functions of any possible additional
labels
are in principle determined as cases of the general expansions we have
been studying. For this reason the present work, although deliberately of a 
generic nature, does indeed confirm the viability of the dynamical symmetry 
approach. 

\pagebreak

\subsection*{Captions of Tables and figures}

{\bf Table \ref{table:dinucprops}}: Table of dinucleoside properties and 
predicted values from 
fits. $ G^{0}_{37}$: Turner free-energy parameters at $37^{\circ}$ in 
kcal mol$^{-1}$ (Xia et al., 1998) ; $R_{f}$: dinucleoside monophosphate 
relative hydrophilicity (Weber and Lacey, 1978); $R_{x}$: dinucleotide 
relative hydrophobicity (Jungck, 1978).

\noindent
{\bf Table \ref{table:aaprops}}:
Table of amino acid properties ($P_{\alpha, \beta}$: conformational parameters 
(Fasman, 1989); $P_{Gr}$: Grantham polarity (Grantham, 1974); $R_{f}$: Relative hydrophilicity (Weber and Lacey, 1978).

\begin{figure}[tbp]
\centering{
\scalebox{0.9}{\includegraphics{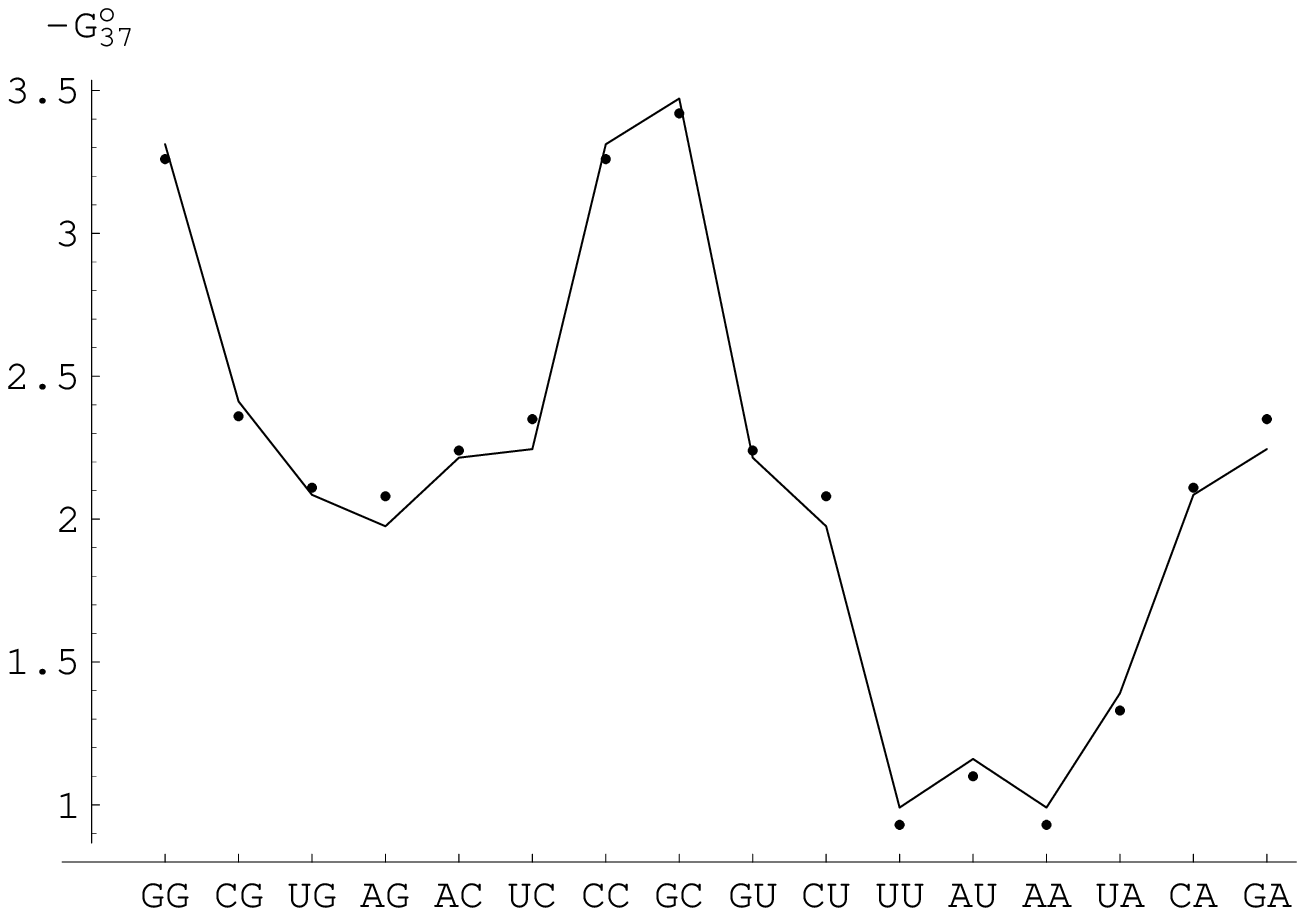}}
}
\caption{}
\protect\label{heatfit}
\end{figure}

\noindent
{\bf Figure \ref{heatfit}}:
Least-squares fit (curve) to the Turner free-energy parameters
(points) at $37^{\circ}$ given by Eq. (\ref{newgfit}). 
Units are in kcal mol$^{-1}$.

\begin{figure}[tbp]
\centering{
\includegraphics{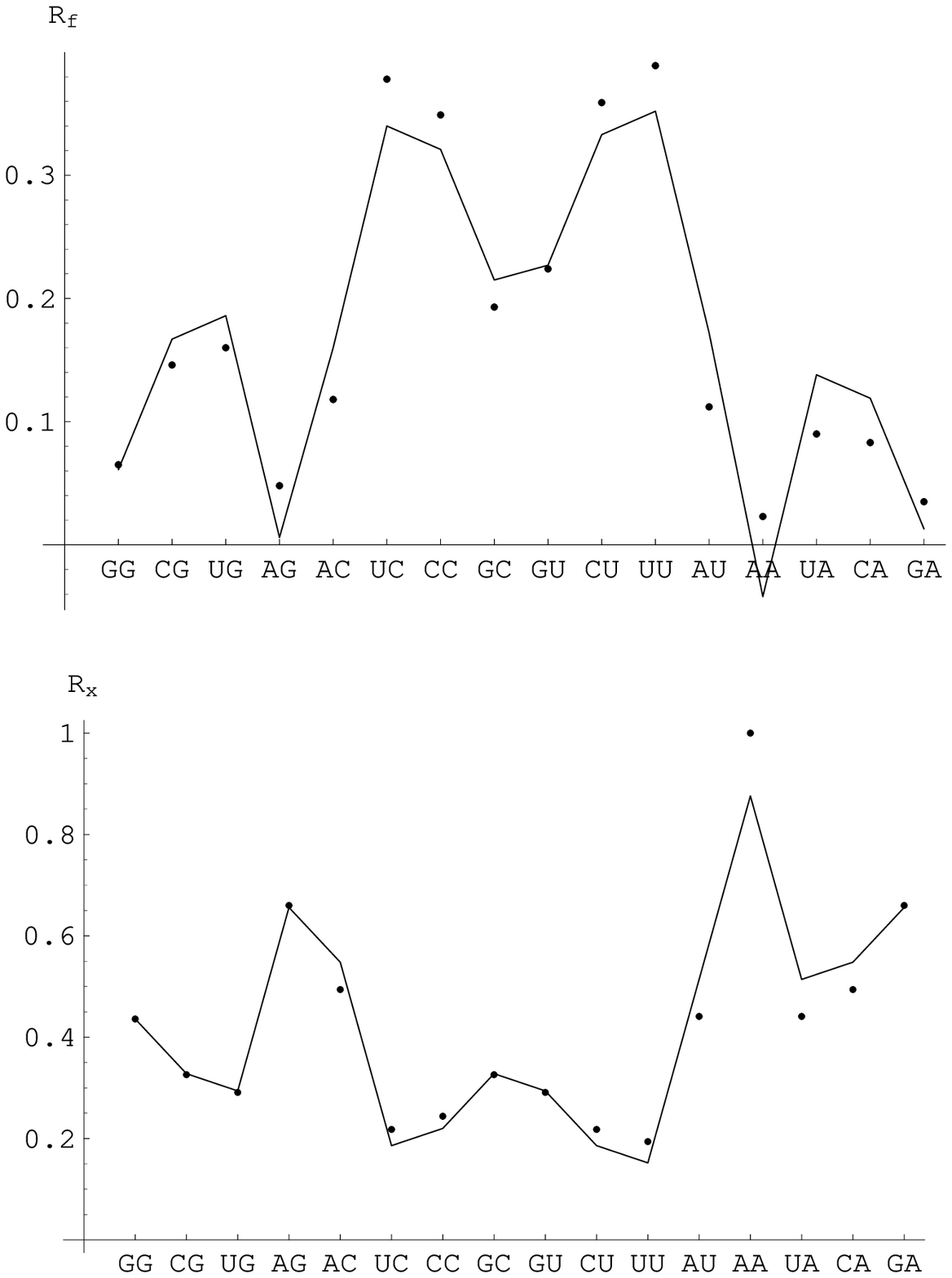}
}
\caption{}
\protect \label{hydrofits}
\end{figure}

\noindent
{\bf Figure \ref{hydrofits}}:
Least-squares fits for dinucleoside $R_{f}$ (upper) and dinucleotide
$R_{x}$ (lower). Points are experimental values while the curves are 
least-squares fits given by Eqs.(\ref{rffit}) and (\ref{rxfit}) respectively.

\begin{figure}[htb]
\centering{
\rotatebox{270}{\resizebox{10cm}{!}{\includegraphics{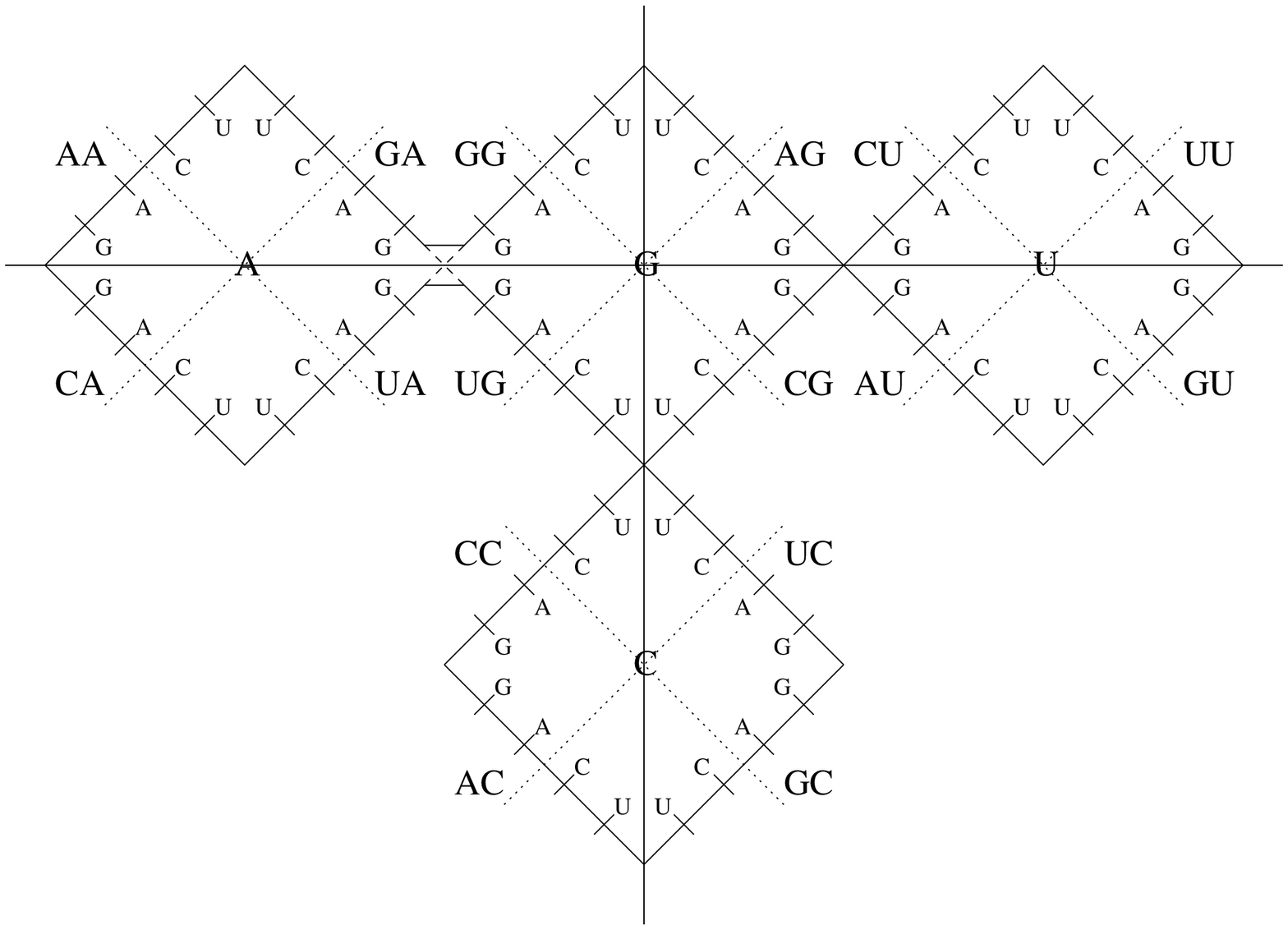}}}
}
\caption{}
\protect \label{diamonds}
\end{figure}

\noindent
{\bf Figure \ref{diamonds}}: 
`Weight diagram' for the genetic code, arising as the
superposition of two projections of the 6-dimensional space of codon
coordinates onto planes corresponding to coordinates for bases of the
first and second codon letters, and an additional one-dimensional
projection along a particular direction in the space of the third
codon base. The orientations of the three projections are chosen
to correspond with the rank ordering of amino acids according to the
one-step mutation rings.

\begin{figure}[htb]
\centering{
\scalebox{0.37}{\includegraphics{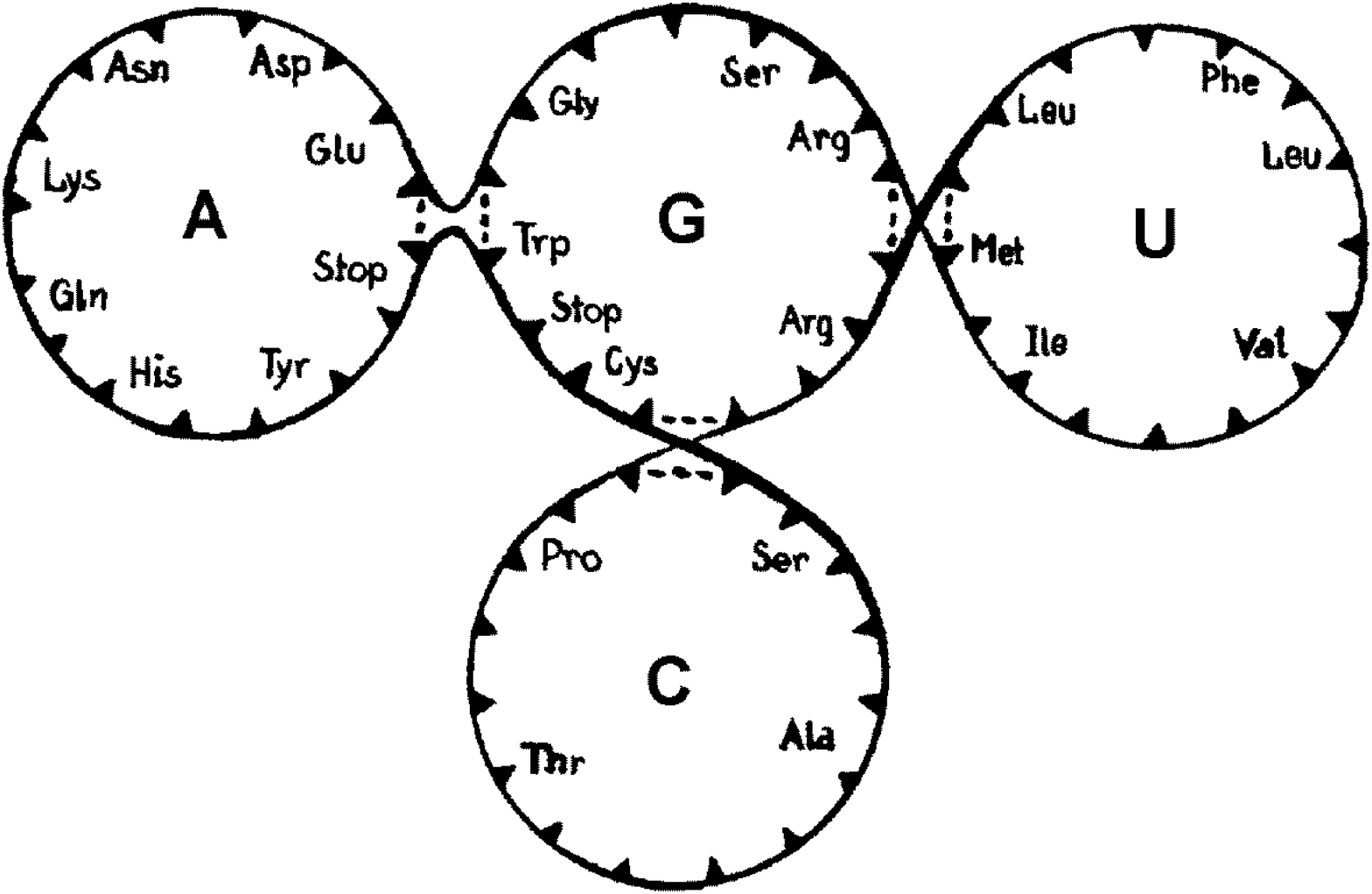}}
}
% \scalebox{0.3}
\caption{}
\protect \label{simrings}
\end{figure}

\noindent
{\bf Figure \ref{simrings}}:
Siemion's interpretation of the weight diagram in terms of
the rank ordering of `one-step mutation rings'. 
Reproduced from Siemion, 1994a.

\noindent
{\bf Figure \ref{palphtrigfits}}:
Least-squares trigonometric fit (dots) to the $P^{\alpha}$ conformational
parameter (small circles) as a function of mutation angle $k$.
Crosses (``pref'') denote the fitted function, Eq.(\ref{PS}), evaluated at 
preferred codon positions. 

\noindent
{\bf Figure \ref{palphtrigfits2}}:
Estimated trigonometric fit (dots) to $P^{\alpha}$ (small circles) as a 
function of $k$. Crosses (``pref'') denote the fitted function, Eq.(\ref{L}), 
evaluated at preferred codon positions. 

\noindent
{\bf Figure \ref{palphmonfits}}:
Polynomial fits (black circles) to the $P^{\alpha}$ conformational parameter (white circles) as a function of the six codon coordinates. The fit is given by 
Eq.(\ref{amon}). Crosses (``modif''): same function, with the $d_{2}^{2}
(d_{3}-m_{3})$ coefficient modified from $-0.075$ to $-0.2$ to enhance 
eight-codon periodicity.

\noindent
{\bf Figure \ref{pbetafit}}:
Least-squares fit (5 parameters) to the $P^{\beta}$ conformational parameter
as a function of the six codon coordinates. 
Small circles: data; dots: least-squares fit given by Eq.(\ref{bmon}).

\noindent
{\bf Figure \ref{aarffit}}:
Least-squares fit (4 parameters) to relative 
hydrophilicity (Weber and Lacey, 1978) as a function of the six codon 
coordinates. Small circles: data; dots: least-squares fit given
by Eq.(\ref{rfmon}).

\noindent
{\bf Figure \ref{grafit}}:
Least-squares fit (2 parameters) to Grantham polarity (Grantham, 1974) as a 
function of the six codon coordinates.
Small circles: data; dots: least-squares fit given by Eq.(\ref{gmon}).

\pagebreak
\begin{table}[tbp]
        \centering
        \caption{}
        \begin{tabular}{|c|c|c|c|c|c|c|}
        \hline
  & $ G^{0}_{37}$ & fit & $R_{f}$ & fit & $R_{x}$ & fit\\
\hline
\hline
GG & -3.26 & -3.312 & 0.065 & 0.0651 & 0.436 & 0.436\\
CG & -2.36 & -2.412 & 0.146 & 0.166 & 0.326 & 0.327\\
UG & -2.11 & -2.085 & 0.16 & 0.185 & 0.291 & 0.293\\
AG & -2.08 & -1.975 & 0.048 & 0.007 & 0.660 & 0.656\\
AC & -2.24 & -2.215 & 0.118 & 0.162 & 0.494 & 0.548\\
UC & -2.35 & -2.245 & 0.378 & 0.341 & 0.218 & 0.186\\
CC & -3.26 & -3.312 & 0.349 & 0.321 & 0.244 & 0.22\\
GC & -3.42 & -3.472 & 0.193 & 0.216 & 0.326 & 0.328\\
GU & -2.24 & -2.175 & 0.224 & 0.227 & 0.291 & 0.293\\
CU & -2.08 & -2.015 & 0.359 & 0.332 & 0.218 & 0.186\\
UU & -0.93 & -0.991 & 0.389 & 0.352 & 0.194 & 0.151\\
AU & -1.10 & -1.161 & 0.112 & 0.173 & 0.441 & 0.514\\
AA & -0.93 & -0.991 & 0.023 & -0.04 & 1     & 0.877\\
UA & -1.33 & -1.391& 0.090 & 0.139 & 0.441 & 0.514\\
CA & -2.11 & -2.085 & 0.083 & 0.119 & 0.494 & 0.548\\
GA & -2.35 & -2.245 & 0.035 & 0.014 & 0.660 & 0.656\\
\hline
\end{tabular}
  \label{table:dinucprops}
\end{table}

\begin{table}[tbp]
	\centering
        \caption{}
	\begin{tabular}{|c|c|c|c|c|}
	\hline
	
$AA$ & $P_{\alpha}$ & $P_{\beta}$ & $P_{Gr}$ & $R_{f}$ \\
\hline
\hline
  ala & 1.38 & 0.79  & 8.09 & 0.89 \\
  arg & 1    & 0.938 & 10.5 & 0.88 \\
  asn & 0.78 & 0.66  & 11.5 & 0.89 \\
  asp & 1.06 & 0.66  & 13   & 0.87 \\ 
  cys & 0.95 & 1.07  & 5.5  & 0.85 \\ 
  gln & 1.12 & 1     & 10.5 & 0.82 \\
  glu & 1.43 & 0.509 & 12.2 & 0.84 \\
  gly & 0.629 & 0.869 & 9   & 0.92 \\
  his & 1.12 & 0.828 & 10.4 & 0.83 \\
  ile & 0.99 & 1.57  & 5.2  & 0.76 \\
  leu & 1.3 & 1.16   & 4.9  & 0.73 \\ 
  lys & 1.20 & 0.729 & 11.3 & 0.97 \\
  met & 1.32 & 1.01  & 5.7  & 0.74 \\
  phe & 1.11 & 1.22  & 5.2  & 0.52 \\
  pro & 0.55 & 0.62  & 8    & 0.82 \\
  ser & 0.719 & 0.938 & 9.19 & 0.96\\
  thr & 0.78 & 1.33 &  8.59 & 0.92 \\
  trp & 1.03 & 1.23 &  5.4 & 0.2 \\
  tyr & 0.729 & 1.31 & 6.2 & 0.49 \\ 
  val & 0.969 & 1.63 & 5.9 & 0.85 \\

		\hline
\end{tabular}
	\label{table:aaprops}
\end{table}

\vspace{8cm}

\begin{figure}[htb]
\centering{
\scalebox{1.4}{\input{biosys-fig5}}
} 
%\scalebox{0.3}
\caption{}
\protect \label{palphtrigfits}
\end{figure}

\begin{figure}[htb]
\centering{
\scalebox{1.4}{\input{biosys-fig6}}
}
\caption{}
\protect \label{palphtrigfits2}
\end{figure}

\begin{figure}[htb]
\centering{
\scalebox{1.4}{\input{biosys-fig7}}
}
\caption{}
\protect \label{palphmonfits}
\end{figure}

\begin{figure}[htb]
\centering{
\scalebox{1.4}{\input{biosys-fig8}}
}
\caption{}
\protect \label{pbetafit}
\end{figure}

\begin{figure}[htb]
\centering{
\scalebox{1.4}{\input{biosys-fig9}}
}
\caption{}
\protect \label{aarffit}
\end{figure}

\begin{figure}[htb]
\centering{
\scalebox{1.4}{\input{biosys-fig10}}
}
\caption{}
\protect \label{grafit}
\end{figure}


\begin{thebibliography}{99}

\bibitem{us1997} Bashford, J.D., Tsohantjis, I. and Jarvis, P.D., 1997.
Codon and nucleotide assignments in a superymmetric model of the genetic code.
Phys. Lett. A 233, 481-488.

\bibitem{us1998} Bashford, J.D., Tsohantjis, I. and Jarvis, P.D., 1999.
A supersymmetric model for the evolution of the genetic code.
Proc. Natl. Acad. Sci. USA 95, 987-992.

\bibitem{BertmanJungck1989} Bertman, M.O. and Jungck,J.R., 1979.
Group graph of the genetic code. J. Heredity 70, 379-384.

\bibitem{ChouFasman1975} Chou, P.Y. and Fasman, G.D., 1974.
Conformational parameters for amino acids in helical, $\beta$-sheet and 
random coil region calulated from proteins. Biochemistry 13, 211-222.

\bibitem{Fasman1989} Fasman, G.D., 1989. The development of prediction of 
protein structure, in: Prediction of Protein Structure
and the Principles of Protein Conformation, G.D. Fasman(ed.) (Plenum, New York) pp. 193-316.

\bibitem{Hornos1997} Forger, M., Hornos, Y. and Hornos, J., 1997.
Global aspects in the algebraic approach to the genetic code. Phys. Rev. E 56, 7078-7082.

\bibitem{Sachse1998} Forger, M. and Sachse, S., 1999.
An orthosymplectic symmetry for the genetic code, in: Group22: Proceedings of 
the XII International Colloquium on Group Theoretical Methods in Physics, 
S.P. Corney, R. Delbourgo and P.D. Jarvis, (Eds.), (International Press, Boston) pp. 147-151.

\bibitem{SorbaSciarrino1998} Frappat, L., Sciarrino, A. and Sorba, P., 1998. 
A crystal basis for the genetic code. Phys. Lett. A 250, 214-221.

\bibitem{SorbaSciarrino2000} Frappat, L., Sciarrino, A. and Sorba, P., 2000. 
Crystalizing the genetic code. physics/0003037

\bibitem{Garel1973} Garel, J.P., Filliol, D. and Mandel, P., 1973.
Coefficients de partage d'aminoacides, nucl\'{e}obases, nucl\'{e}osides et nucl\'{e}otides dans un syst\`{e}me solvant salin. J. Chromatog. 78, 381-391.

\bibitem{GoodmanMoore1977} Goodman, M. and Moore, G.W., 1977.
Use of Chou-Fasman amino acid conformational parameters to analyze the
organization of the genetic code and to construct protein genealogies.
J. Mol. Evol. 10, 7-47.

\bibitem{Grantham1974} Grantham, R., 1974. Amino acid difference formula to 
help explain protein evolution. Science 185, 862-864.

\bibitem{HornosHornos1993} Hornos, J. and Hornos, Y., 1993. 
Algebraic model for the evolution of the genetic code. Phys. Rev. Lett. 71, 
4401-4404.

\bibitem{usICGTMP1998} Jarvis, P.D. and Bashford, J.D., 1999.
Systematics of the genetic code and anticode: history, supersymmetry, 
degeneracy and periodicity, in: Group22: Proceedings of 
the XII International Colloquium on Group Theoretical Methods in Physics, 
S.P. Corney, R. Delbourgo and P.D. Jarvis, (Eds.), (International Press, 
Boston) pp. 143-146.

\bibitem{Jukes1973} Jukes, T.H., 1973. Arginine as an evolutionary intruder 
into protein synthesis. Biochem. Biophys. Res. Comm. 53, 709-714.

\bibitem{Jungck1978} Jungck, J.R., 1978. The genetic code as a periodic table.
J. Mol. Evol. 11, 211-224.

\bibitem{LaceyMullins1983} Lacey, J.C. Jnr. and Mullins, D.W., 1983.
Model for the coevolution ofthe genetic code and the process of protein 
synthesis. Origins of Life 13, 3-42.

\bibitem{Zuker1999} Mathews, D.H., Sabina, J., Zuker, M. and Turner, D.H., 
1999. Expanded sequence dependence of thermodynamic parameters improves 
prediction of RNA secondary structure. J. Mol. Biol. 288, 911-940. 

\bibitem{JukesReview1991} Osawa, S., Jukes, T.H., Watanabe, K. and Muto, A., 
1992. Recent evidence for evolution of the genetic code. Microbiol. Rev. 56, 
229-264.

\bibitem{SchlesingerWK1998} Schlesinger, M., Kent, R.D. and Wybourne B.G., 
1997. Group theory and the meaning of life?, in : Proc. 4th International
Summer School in Theoretical Physics (Singapore: World Scientific, 1997)
pp. 263-282. 

\bibitem{SchlesingerWK1999}
Schlesinger, M. and  Kent, R.D., 1999.
On algebraic approaches to the genetic code, in: Group22: Proceedings of 
the XII International Colloquium on Group Theoretical Methods in Physics, 
S.P. Corney, R. Delbourgo and P.D. Jarvis, (Eds.), (International Press, 
Boston) pp. 152-159.

\bibitem{Turner1995} Serra, M.J. and Turner, D.H., 1995.
Predicting thermodynamic properties of RNA. Meth. Enzymol. 259, 243-261.

\bibitem{Turner1986} Frier, S.M., Kierzek R., Jaeger, J.A., Sugimoto, N.,
Caruthers, M.H., Neilson, T. and Turner, D.H., 1986.
Improved free-energy parameters for predictions of RNA duplex stability.
Proc. Natl. Acad. Sci. USA 83, 9373-9377.

\bibitem{Siemion1992} Siemion, I.Z. and Stefanowicz, P., 1992.
Periodical changes of amino acid reactivity within the genetic code.          
BioSystems 27, 77-84.

\bibitem{Siemion1994a} Siemion, I.Z., 1994a. The regularity of changes of
Chou-Fasman parameters within the genetic code. BioSystems 32, 25-35.

\bibitem{Siemion1994b} Siemion, I.Z., 1994b. Compositional frequencies of 
amino acids in the proteins and the genetic code. BioSystems 32, 163-170.

\bibitem{SiemionSiemion1994} Siemion, I.Z. and Siemion, P.J., 1994. The 
informational context of the third base in amino acid codons. BioSystems 33,
139-148.

\bibitem{Siemion1995} Siemion, I.Z., Siemion, P.J. and Krajewski, K., 1995. 
Chou-Fasman conformational amino acid parameters and the genetic code. 
BioSystems 36, 231-238.

\bibitem{Siemion1996} Siemion, I.Z. and P Stefanowicz, 1996. The presumable
place of ornithine in an earlier genetic code. Bull. Polish Acad. Sci. 44, 
63-69.

\bibitem{SjostromWald1985} Sj\"{o}str\"{o}m, M. and Wold, S., 1985.
A multivariate study of the relationship between the genetic code and the
physical-chemical properties of amino acids. J. Mol. Evol. 22, 272-277.

\bibitem{TaylorCoates1989} Taylor, F.J.R. and Coates, D., 1989.
The code within codons. BioSystems 22, 177-187.

\bibitem{Volkenstein1966} Volkenstein, M.V., 1966. The genetic coding of the
protein structure. Biochim. Biohys. Acta 119, 421-424.

\bibitem{WeberLacey1978} Weber, A.L. and Lacey, J.C. Jnr., 1978.
Genetic code correlations: amino acids and their anticodon nucleotides.
J. Mol. Evol. 11, 199-210.

\bibitem{Woese1966} Woese, C.R., Dugre, D.H., Kando, M. and Saxinger, W.C., 
1966. On the fundamental nature and evolution of the genetic code. 
Cold Spring Harbour Symp. Quant. Biol. 31, 723-736.

\bibitem{Xia1998} Xia, T., SantaLucia, J. Jnr., Burkard, M.E.,
Kierzek, R., Schroeder, S.J., Jiao, X., Cox, C., and Turner, D.H., 1998.
Parameters for an expanded nearest-neighbour model for formation of RNA
duplexes with Watson-Crick pairs. Biochemistry 37, 14719-14735.



\end{thebibliography}
\end{document}